# Nanomolding of Metastable Mo$_4$P$_3$

Mehrdad T Kiani[a], Quynh P Sam[a], Gangtae Jin[a], Betül Pamuk[b], Hyeuk Jin Han[a,c], James L. Hart[a], J. R. Stauff[d], Judy J Cha[a]

a) Department of Materials Science and Engineering, Cornell University, Ithaca, NY, 14853, USA
b) School of Applied and Engineering Physics, Cornell University, Ithaca, NY 14853, USA
c) Department of Environment and Energy Engineering, Sungshin Women's University, Seoul 01133, South Korea
d) Department of Mechanical Engineering and Materials Science, Yale University, New Haven, CT, 06520, USA

**Abstract**

Reduced dimensionality leads to emergent phenomena in quantum materials and there is a need for accelerated materials discovery of nanoscale quantum materials in reduced dimensions. Thermomechanical nanomolding is a rapid synthesis method that produces high quality single-crystalline quantum nanowires with controlled dimensions over wafer-scale sizes. Herein, we apply nanomolding to fabricate nanowires from bulk feedstock of MoP, a triple-point topological metal with extremely high conductivity that is promising for low-resistance interconnects. Surprisingly, we obtained single-crystalline Mo$_4$P$_3$ nanowires, a metastable phase at room temperature in atmospheric pressure. We thus demonstrate nanomolding can create metastable phases inaccessible by other nanomaterial syntheses and can explore a previously inaccessible synthesis space at high temperatures and pressures. Furthermore, our results suggest that the current understanding of interfacial solid diffusion for nanomolding is incomplete, providing opportunities to explore solid-state diffusion at high-pressure and high-temperature regimes in confined dimensions.

**Introduction**

One-dimensional (1D) material systems display emergent phenomena due to reduced dimensionality and nanoscale confinement not present in higher dimensions, such as dislocation starvation in nanopillars[1], deviations in crystallization in metallic glasses[2], and ballistic transport in 1D van der Waals crystals[3]. In the context of quantum materials, realizing topological superconductors for probing Majorana bound states[4], maximizing topological surface states for low dissipation microelectronics[5] or catalysis[6], and developing Josephson junctions with unusual current-phase relations[7] all rely on pristine 1D nanowire systems. Given the large predicted number of topological quantum materials[8], there are limited nanofabrication techniques that allow for simultaneous morphological, crystallographic, and structural control of 1D nanowires over wafer-scale distances. Traditional bottom up fabrication techniques such as chemical vapor deposition or molecular beam epitaxy require extensive optimization and yet control of size and defect structure is limited, while top-down lithographic approaches can achieve the desired size but lack any control of defect structure and are severely limited with respect to material choice[9].

In 2019, Liu *et al.* introduced the scalable fabrication method of thermomechanical nanomolding (TMNM), where a bulk polycrystalline feedstock material is extruded through a nanoporous mold at elevated temperatures (approximately $0.5T_m$, where $T_m$ is the melting point) and pressures (>100 MPa)[10]. This process results in formation of single-crystalline, defect-free nanowires with high aspect ratios[11]. The mold is etched away using a strong acid or base, and the molded nanowires are separated from the bulk feedstock via sonication. TMNM holds several key advantages over traditional synthesis methods, including the ability to produce single-crystalline nanostructures from a polycrystalline feedstock due to re-orientation of the grains as they are pressed through the mold[12]. Another major advantage of TMNM is its versatility as the library of successfully nanomolded materials has expanded to include crystalline metals[13], solid solutions[10], and ordered phases[14].

The use of TMNM for 1D quantum materials has been unexplored. One particular issue is that many quantum materials of interest possess covalent bonding, such as metal-phosphides[15–18]. Molybdenum-phosphides such as $MoP$[16,19], $MoP_2$[20,17], and $MoP_4$[15] exhibit unique quantum transport effects arising from topologically protected surface states and MoP, a triple point topological metal[21], is a promising alternative to Cu due to its high carrier density, high electron mobility, and low bulk resistivity. In Mo-P compounds, metallic bonding exists between Mo atoms; however, the shortest atomic bond distances are between covalently bonded Mo and P atoms[22]. Previous TMNM studies focused on metals or intermetallics with largely metallic bonding[13,14], which can make bond rearrangement during solid-state diffusion easier while certain covalently bonded materials such as Si are not suitable for TMNM given their extremely low bulk and surface diffusivities[14].

We herein report the nanomolding of single-crystalline nanowires from a polycrystalline MoP feedstock. Surprisingly, crystallographic and compositional analysis shows the nanowires to be $Mo_4P_3$ instead of MoP. $Mo_4P_3$ is a metastable phase at room temperature and ambient pressure with little experimental characterization. Density functional theory (DFT) band structure calculations and resistivity measurements indicate that $Mo_4P_3$ is a non-topological metal with resistivity values comparable to other molybdenum phosphide compounds. These results demonstrate that high temperatures and pressures, along with the interfacial energy between the feedstock and mold materials in TMNM, can provide a novel pathway to nanofabrication of metastable phases.

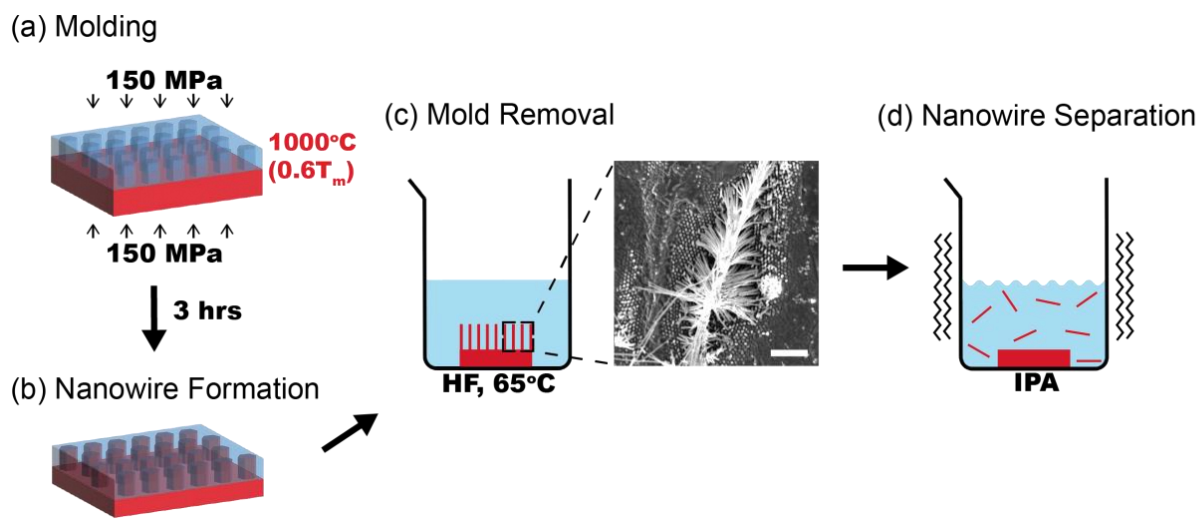

**Figure 1.** Schematic of the TMNM process. (a) Molding of polycrystalline MoP feedstock (red) using an anodic aluminum oxide mold (blue) at high temperature and pressure. (b) Formation of single-crystalline nanowires after a mold period of 3 hrs. (c) Mold removal using hydrofluoric acid (HF) and scanning electron microscope (SEM) image of nanowires molded from MoP feedstock (scale bar = 1µm). (d) Separation of nanowires from feedstock via sonication in isopropyl alcohol (IPA).

**Results**

Bulk polycrystalline MoP, which has been polished to a mirror finish, is used as the initial feedstock material (Figure S1). TMNM is performed at 1000°C and 150 MPa for 3 hours using an anodic aluminum oxide (AAO) mold with 40 nm pores (Figure 1a, b) in an argon environment. Etching of the mold in HF and sonication lead to dispersed nanowires in solution (Figure 1c, d), which are dropcast onto TEM grids or SiO$_2$ substrates for characterization and electrical measurements, respectively. Details of the TMNM can be found in Supporting Information.

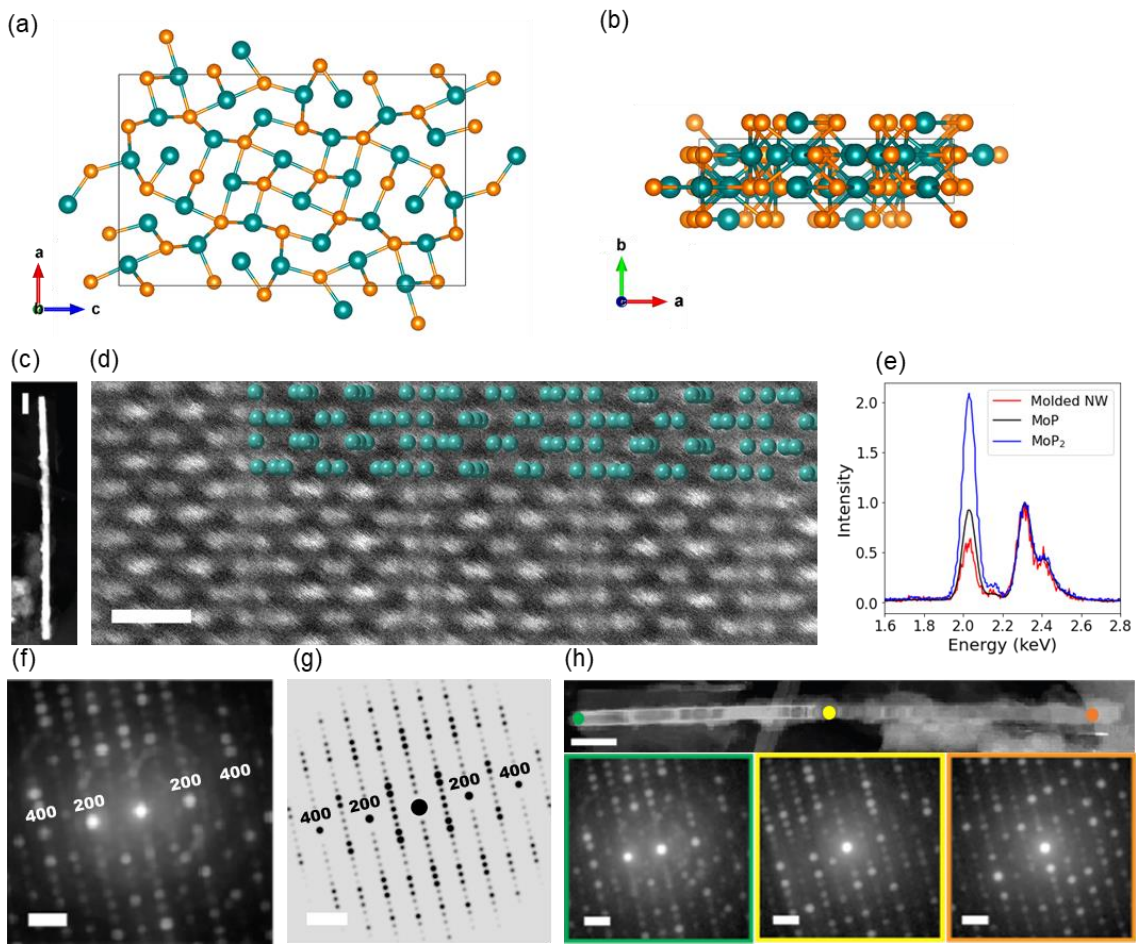

**Figure 2.** Atomic characterization of molded nanowires. (a, b) Schematics of Mo$_4$P$_3$ crystal structure (Mo atoms: teal, P atoms: orange). (c) Low-magnification image of nanowire (scale bar = 0.1 µm). (d) STEM image of nanowire (scale bar = 5 Å). (e) EDX spectra of molded nanowire, reference MoP, and MoP$_2$ bulk single crystals. (f) Integrated diffraction pattern over 6x6 pixel area of nanowire from 4D STEM (scale bar = 0.5 Å$^{-1}$). (g) Simulated diffraction pattern of Mo$_4$P$_3$ (scale bar = 0.5 Å$^{-1}$). (h) Diffraction patterns (scale bar = 0.5 Å$^{-1}$) from three points along the length of the nanowire (scale bar = 0.1 µm).

Scanning transmission electron microscopy (STEM) is used to characterize the molded nanowires. The average diameter of the nanowires is 33.8±1.9 nm. High-angle annular dark field (HAADF) STEM images reveal that the nanowires have a different atomic structure from their bulk MoP feedstock (Figure S1). The high-resolution STEM image in Figure 2d shows the atomic structure of the nanowires is not the hexagonal crystal structure of MoP or MoP$_2$, shown in Figure S2. From crystallographic analysis, we determine the phase of the nanowires to be orthorhombic

Mo₄P₃ (Figure 2a, b), a high-temperature compound stable between 700°C-1580°C at ambient pressure[23]. As shown by the overlay in Figure 2d, the atomic positions in the STEM image closely match those of the Mo atoms in Mo₄P₃, with the growth direction along the $b$ axis. The lattice constant perpendicular to the growth direction of the nanowire is measured to be 21.3 Å, which nearly matches the Mo₄P₃ $c$ lattice constant of 20.5 Å[24]. The 3.9% discrepancy can be attributed to tensile strain along the curvature of the nanowire. Using the fast fourier transform (FFT) of the STEM image, shown in Figure S3, we calculate the lattice constant along the growth direction of the nanowire to be 3.2 Å, in close agreement with the $b$ lattice constant of Mo₄P₃ of 3.16 Å[24]. Figure 2e shows the normalized energy dispersive X-ray spectroscopy (EDX) spectrum of the molded nanowires in comparison with two reference spectra measured from bulk single crystals of MoP and MoP₂. Using the reference spectra, the Mo to P ratio of the nanowires is calculated to be nearly 4:3 (Table S1). STEM-EDX maps (Figure S4) show that the distribution of Mo and P is uniform over the entire area of the nanowires. No Mo₄P₃ nanowire had noticeable oxide formation; however, some had residual AAO mold.

Diffraction data is collected via 4D STEM scan over the nanowire shown in Figure 2c to confirm its single-crystalline nature. Figure 2f shows a diffraction pattern generated by summing up the 4D STEM data over an area of 726 nm² on the nanowire outlined in Figure S5. The diffraction pattern shows that the nanowire is single crystalline and a close qualitative match with the simulated diffraction pattern of Mo₄P₃ shown in Figure 2g. The area of integration for the diffraction pattern was limited to avoid shifts in the diffraction spots arising from changes in nanowire orientation on the TEM grid due to bending, which would distort the final diffraction pattern. In Figure 2h, single diffraction patterns taken from different points along the wire length confirm that the nanowire is single crystalline, unlike the starting bulk polycrystalline MoP (Figure S1). While there are changes in the intensity of the diffraction spots due to varying orientation along the length of the nanowire, the overall diffraction spot positions remain consistent owing to the lack of grains.

There are limited experimental studies on transport properties of bulk Mo₄P₃ since it is a metastable phase not found at room temperature. Shirotani *et al.* found that bulk Mo₄P₃ prepared under high pressure exhibits superconductivity at 3 K[22] but they do not measure room temperature resistivity. The electronic band structure calculated from DFT (Figure 3a) reveals that Mo₄P₃ is metallic. Projected band structures and density of states show that Mo₄P₃ is not topological with mainly the Mo *d* orbitals contributing to bands near the Fermi level (Figure 3a and Figure S6). The total density of states is comparable with other Mo-P intermetallic compounds[25]. The growth direction of the molded nanowires and subsequently, the direction of current flow, is along the *b* axis, which corresponds to Γ-*Y* direction in the band structure (Figure 3b). To measure resistivity, we fabricated four-point probes on $Mo_4P_3$ nanowires deposited on $SiO_2$ using e-beam deposited Ti/Au electrodes (Figure 3c). From four-point probe measurements, the resistivity of the $Mo_4P_3$ nanowires is measured to be 39.7±3.4 µΩ·cm at 300 K (Figure 3d). Since no experimental resistivity values exist for bulk or nanostructure $Mo_4P_3$, we compare the resistivity to *ab initio* calculations, which report a range from 33 – 334 µΩ·cm (Table S2, S3)[26]. Since the resistivity values of $Mo_4P_3$ nanowires matches the predicted bulk value, we hypothesize that surface electron scattering is minimal even though the sample is not topological. The resistivity value of the $Mo_4P_3$ nanowires measured is higher than that of MoP nanowires[27], 12 µΩ·cm, for the equivalent cross-sectional area but is comparable to single crystal 60 nm diameter $MoP_2$ nanowires[20], 32 µΩ·cm.

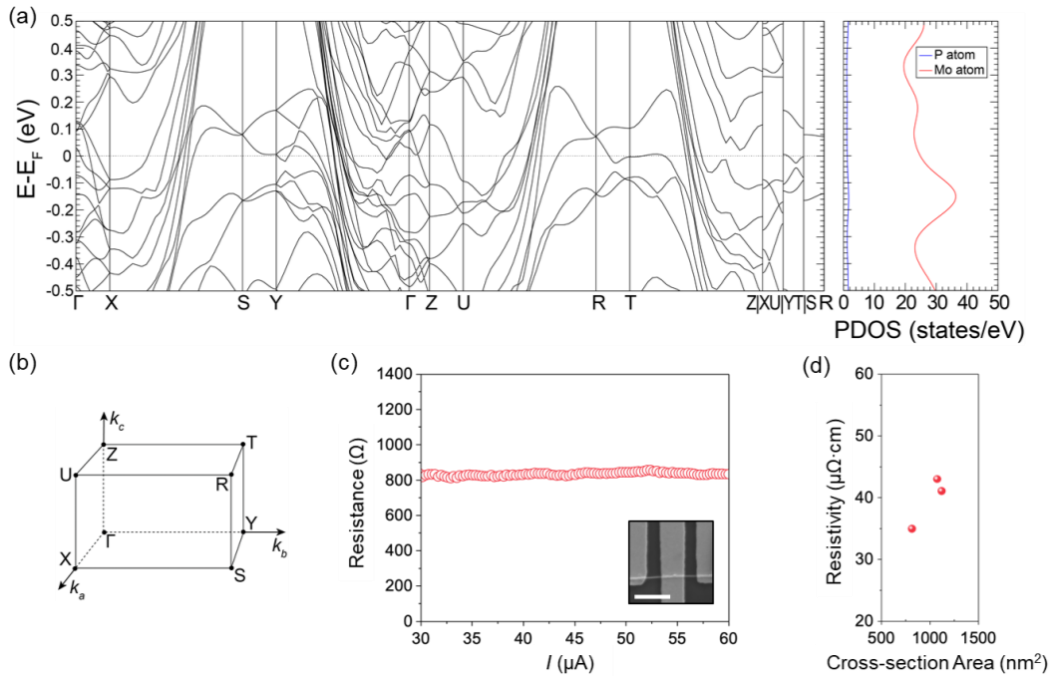

**Figure 3.** (a) Predicted band structure and partial density of states near Fermi level for $Mo_4P_3$. (b) High symmetry labels for Brillouin zone of $Mo_4P_3$ (c) Resistance versus current of $Mo_4P_3$ during four-point probe measurement. Inset is SEM image of four-point probe. (d) Resistivity of various nanowires with differing cross-sectional area.

Previous TMNM focused on ordered phases with largely metallic bonding between atoms. Here, we show that nanomolding can be successful on metal phosphides, which possess largely covalent bonding. A key distinction of the present work is that, while the bulk feedstock remained MoP, change in crystal structure and loss of phosphorus occurred during nanomolding to stabilize and produce metastable $Mo_4P_3$ whose composition is different from the feedstock, distinct from previous experiments where chemical composition matched the bulk[14]. We propose two potential mechanisms that could lead to loss of phosphorus in the sample. Due to the high temperatures involved during molding and the high vapor pressure of phosphorus compared to molybdenum[28], the nanowires could slowly lose phosphorus leading to a change in chemical composition and crystal structure. In addition, phosphorus could be diffusing into the AAO mold. Alumina has a strong binding affinity for phosphates[29] and lateral diffusion of phosphorus into the mold during growth could lead to a change in composition.

In conclusion, we show that TMNM is a viable nanofabrication strategy for 1D quantum materials. Using a bulk MoP feedstock, we nanomold $Mo_4P_3$ nanowires and confirm their single crystalline, defect-free structure. To date, this is the only nanofabrication technique for single crystal 1D nanowires for $Mo_4P_3$. DFT calculations showed a non-topological character for $Mo_4P_3$, and resistivity values of nanowires are comparable to other molybdenum phosphides and close to predicted bulk resistivity, indicating that surface scattering is minimal. Since TMNM is material agnostic, the approach elucidated here can be extended to fabricate other metastable metal phosphides or covalently bonded systems.

**Acknowledgements**

We acknowledge the support of the National Science Foundation Metals and Metallic Nanostructures program for TMNM (DMR 2240957), Semiconductor Research Corporation for transport measurements (nCORE IMPACT), and Gordon and Betty Moore Foundation for 4D STEM experiments (EPiQS Synthesis Investigator Award). Q.S. acknowledges support from the NSF Graduate Research Fellowship Program. Characterization of the nanowires and DFT calculations were made possible by the Platform for the Accelerated Realization, Analysis, and Discovery of Interface Materials (PARADIM) supported by NSF Cooperative Agreement No. DMR-2039380, Cornell Center for Materials Research (CCMR) Shared Facilities supported by the NSF MRSEC program (DMR-1719875), and Cornell Nanoscale Facility (CNF) supported by NSF grant No. NNCI-2025233. TEM sample preparation was performed on a Helios FIB at the CCMR facilities supported by NSF (DMR-1539918).